\documentclass[aps,prd,twocolumn,showpacs,superscriptaddress,letterpaper,asmmath,amssymb]{revtex4}

\usepackage{graphicx}
\usepackage{dcolumn}
\usepackage{bm}
\usepackage{feynmf}

\RequirePackage{xspace}
\usepackage{relsize}

\def\W      {\ensuremath{W}\xspace}

\def\t     {\ensuremath{t}\xspace}

\def\Kbar  {\kern 0.2em\overline{\kern -0.2em K}{}\xspace}

\def\Bbar    {\kern 0.18em\overline{\kern -0.18em B}{}\xspace}

\def\tprime  {\ensuremath{t^{\prime}}\xspace}
\def\bprime  {\ensuremath{b^{\prime}}\xspace}

\def\missET {{\not\!\! E_T}}
\def\Qbar    {\kern 0.08em\overline{\kern -0.08em Q}{}\xspace}

\def\lfj {\ensuremath{\ell+4j}\xspace}
\def\lljbme {\ensuremath{\ell^{\pm}\ell^{\pm}jb\missET}\xspace}

\def\invfb   {\ensuremath{\mbox{\,fb}^{-1}}\xspace}
\newcommand{\mev}{\ensuremath{\mathrm{\,Me\kern -0.1em V}}\xspace}
\newcommand{\mevc}{\ensuremath{{\mathrm{\,Me\kern -0.1em V\!/}c}}\xspace}
\newcommand{\mevcc}{\ensuremath{{\mathrm{\,Me\kern -0.1em V\!/}c^2}}\xspace}
\newcommand{\gev}{\ensuremath{\mathrm{\,Ge\kern -0.1em V}}\xspace}
\newcommand{\gevc}{\ensuremath{{\mathrm{\,Ge\kern -0.1em V\!/}c}}\xspace}
\newcommand{\gevcnospace}{\ensuremath{{\mathrm{\,Ge\kern -0.1em V\!/}c}}}
\newcommand{\gevcc}{\ensuremath{{\mathrm{\,Ge\kern -0.1em V\!/}c^2}}\xspace}
\newcommand{\bea}{\begin{eqnarray}}
\newcommand{\eea}{\end{eqnarray}}

\def\missET {{\not\!\! E_T}}

\begin{document}
\bibliographystyle{apsrev}

\title{Direct Mass Limits for Chiral Fourth-Generation Quarks in All Mixing Scenarios}

\author{Christian J. Flacco}
\email{cflacco@mac.com}
\author{Daniel Whiteson}
\email{daniel@uci.edu}
\author{Tim M.P. Tait}
\email{tmptait@gmail.com}
\affiliation{Department of Physics and Astronomy, University of California, %
Irvine, CA 92697, USA}
\author{Shaouly Bar-Shalom}
\email{shaouly@physics.technion.ac.il}
\affiliation{Physics Department, Technion-Institute of Technology, Haifa 32000, Israel}

\begin{abstract}
Present limits on chiral
fourth-generation quark masses $m_{\bprime}$ and $m_{\tprime}$  are broadly generalized and strengthened by combining both $t^\prime$ and $b^\prime$ decays and considering the full range of $t^\prime$ and $b^\prime$ flavor-mixing scenarios (with the lighter generations). Various characteristic mass-splitting choices are considered. With $m_{\tprime} > m_{\bprime}$ we find that CDF limits on the $b^\prime$ mass vary by no more than 10-20\% with any choice of flavor-mixing, while for the $t^\prime$ mass, we typically find stronger bounds, in some cases up to
$m_{\tprime} > 430$ GeV. For $m_{\bprime} > m_{\tprime}$ we find
$m_{\bprime} > 380 - 430$ GeV,
depending on the flavor-mixing and the size of the $m_{\tprime} - m_{\bprime}$ mass splitting.

\end{abstract}



\maketitle

\date{\today}


Despite many constraints by precision electroweak data, the number of simple Standard
Model (SM) quark generations has not been definitively established. Recent searches by the CDF
Collaboration placing lower limits on the masses of such objects \cite{CDFb,CDFt} have been
construed to leave open significant portions of the theoretical landscape
that include a light fourth generation~\cite{reviews,DEWSB,hung,hashimoto1,hashimoto2}.  In this letter, we offer a treatment of the published data
that fully probes the robustness of these limits and applies them to
heavy quark decay modes for which there are no present direct limits.
By considering all scenarios of mixing among four SM generations, we demonstrate that present data
exclude a chiral fourth generation quark with a mass up to nearly 300 GeV, independent of the flavor-mixing assumptions.
In so doing we establish a methodology for interpretation of future mass limits.

Recent searches for direct production of
fourth generation quarks, denoted $t'$ and $b'$
for the up- and down-type,
found $m_{t'}>335$ GeV \cite{CDFt} and $m_{b'}>338$ GeV \cite{CDFb}, assuming
$\mathcal{B}(t'\rightarrow W\{q=d,s,b\})=100$\% and
$\mathcal{B}(b'\rightarrow Wt)=100$\% respectively.
Each assumes that an individual contribution from one flavor of fourth-generation quark comprises
the entire signal.
These searches have been interpreted under the assumptions of
$m_{t'} - m_{b'} < M_W$ and negligible mixing of
the $(t',b')$ states with the two lightest quark generations.
To account for electroweak precision data \cite{EWPD}
and flavor data \cite{FP} such conditions are typically required for 
SM extensions with four quark generations
and one Higgs
doublet; however, more exotic scenarios have also been 
proposed \cite{hung,hashimoto1,hashimoto2,He:2001tp} .

This subtle theoretical landscape demands a broader view: there is no
uniquely interesting set of assumptions under which
experimental data should be interpreted.
From the experimental view point, choosing a simple set of assumptions allows straightforward,
if narrow, interpretation of results. Yet these interpretations may be
extended beyond the narrow mixing assumptions, and even applied to
previously unexplored decay modes.
In this article
we consider the possible 4th generation mass spectrum and
flavor-mixing space broadly to show that data from the
CDF searches set direct
limits for arbitrary
mixing values and for both cases of $|m_{t'} - m_{b'}| > M_W$.
In particular, we present results for the cases
$m_{t'}>m_{b'}$ and $m_{b'}>m_{t'}$, and both
modest ($50~\textrm{GeV}$) and large ($100~\textrm{GeV}$)
mass splitting.

CDF analyzed two independent samples:
(1) a collection of events containing a single lepton and at least four jets,
denoted as the \lfj sample \cite{CDFt},
and (2) a collection of events
containing two same-charge leptons, two jets (one with a flavor tag) and
evidence of neutrinos (missing
transverse energy), denoted as the  \lljbme sample \cite{CDFb}. We describe them here.


In 4.6~\invfb of data, CDF searched for $t'\rightarrow W\{q=d,s,b\}$ decays in the mode
\[ p \bar p \rightarrow  t'\bar{t'}\rightarrow (W\rightarrow l\nu) q(W\rightarrow qq')q \]
by requiring a single lepton and at
least four jets \cite{CDFt}.
The data were analyzed by reconstructing the invariant
mass of the candidate $t'$ and measuring the total energy in
the event. The event selection used the four jets of highest transverse energy in the event, but
did not require a flavor signature ($b$-tag) on any of the jets, making it generally
sensitive to $t'\rightarrow W\{q=d,s,b\}$.  Assuming $\mathcal{B}(t'\rightarrow
W\{q=d,s,b\})=100\%$, CDF found $m_t' > 335$ GeV \cite{CDFt}.

The mass reconstruction used minimum-likelihood fitting methods that depend upon the particular
spectrum of final state components.  If, for
example, one half of an event decayed as $t'\rightarrow W(b'
\rightarrow Wq)$, giving a $WWqWq$ topology, it might satisfy
 the $\ell+4j$ selection criteria, but the reconstructed mass
distribution for such events would be significantly modified by the additional
$W$. Thus, the results cannot
be trivially applied to topologies other than $WqWq$.
We therefore apply the $\ell+4j$ results exclusively to $WqWq$
processes.


CDF also searched for $b'\rightarrow Wt$ decays in 2.7\invfb of data, in the
same-charge lepton mode:
\begin{eqnarray}
p \bar p \rightarrow b'\bar{b'} \rightarrow WtW\bar{t} & \rightarrow & WWbWW\bar{b} \nonumber \\
 & \rightarrow & ( \ell^{\pm} \nu)(qq')b(qq')(\ell^{\pm}
\nu)\bar{b} \nonumber
\end{eqnarray}
by requiring two same-charge
leptons, at least two jets (at least one with a $b$-tag), and missing
transverse energy of at least 20 GeV \cite{CDFb}.
Given the small backgrounds, multiple neutrinos and large jet multiplicity in the sample,
CDF did not reconstruct the $b'$ mass, but instead fit the observed jet
multiplicity to signal and background templates generated from simulations. Assuming
$\mathcal{B}(b'\rightarrow Wt)=100\%$, CDF found $m_b' > 338$ GeV \cite{CDFb}.

The $\ell^{\pm}\ell^{\pm}jb\missET$ analysis did not use final-state dependent
fits, thus results are process-independent and may be applied to any process producing
the $\ell^{\pm}\ell^{\pm}jb\missET$ signal. For example, $\tprime\rightarrow\W\bprime\rightarrow
\W\W\t\rightarrow\W\W\W b$ decays would produce a six-$W$, two-$b$ signature,
with higher jet multiplicity and
larger acceptance to the $\ell^{\pm}\ell^{\pm}jb\missET$ sample than
the simple four-$W$, two-$b$ signature. In this analysis, we therefore apply
the $\ell^{\pm}\ell^{\pm}jb\missET$ results inclusively to processes resulting in at least
four $W$ bosons and two $b$ quarks.

\begin{figure}[ht]
\includegraphics[width=0.55\linewidth]{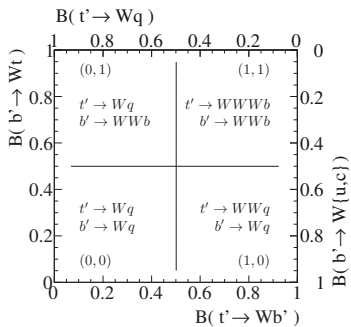}
\caption{ The flavor-mixing intervals overlaid with a table of the processes contributing
to the axis vertices.}
\label{fig:corners}
\end{figure}

The $\ell+4j$ and $\ell^{\pm}\ell^{\pm}jb\missET$ data samples are complementary.
In the case that the fourth generation quarks decay exclusively at tree level through
the charged current electroweak interaction (assured for a chiral
fourth generation),  the two searches
can be minimally understood to probe two corners of a two-dimensional interval
in branching fraction space.
In particular, for the case where the $t'$ is heavier than the $b'$
the topologies of $b'$ and
$t'$ decays are determined by four branching fractions, two of which are independent:
\begin{eqnarray}
\mathcal{B}(t'\rightarrow Wb') & =  &1 - \mathcal{B}(t'\rightarrow W\{q=d,s,b\}) \nonumber \\
\mathcal{B}(b'\rightarrow Wt) & = & 1 - \mathcal{B}(b'\rightarrow W\{q=u,c\})  \nonumber
\end{eqnarray}
as shown in Fig.~\ref{fig:corners}. 
In this representation, the  \lljbme and \lfj analyses probe complementary regions (see Fig.~\ref{fig:corners}).

We consider the implications of the CDF data to various two-flavor
($t'$ and $b'$) scenarios,
characterized by the $\tprime - \bprime$ mass splitting and flavor-mixing rates.
To extend the interpretations of the published results, we use
the relationship among event yield, cross section and acceptance to interpret the
observed yield limits under the varying assumptions. This requires parametrization of
the variation in relative
acceptance due to modification of the signal-source model.

First, we consider the \lljbme sample, interpreted under the mass splitting
assumption $m_{t^\prime} > m_{b^\prime}$.
In the original \lljbme analysis, the event yield was
assumed to come from an individual \bprime . Here, we interpret this yield under a model
with both \tprime and \bprime contributions, where the decay modes of interest are
\begin{eqnarray}
b' & \rightarrow & Wt \rightarrow WWb  \nonumber \\
t' & \rightarrow & Wb'\rightarrow WWt \rightarrow WWWb \nonumber
\end{eqnarray}
which corresponds to the boundary case (1,1) in branching fraction space.
The $t' \rightarrow WWWb$ mode has no prior
direct limit despite having a similar signature with larger acceptance in the
$\ell^{\pm}\ell^{\pm}jb\missET$ dataset due to the two additional
$W$s in the intermediate decay chain. Indeed, if both fourth-generation
quarks exist, we expect to select both
modes in the $\ell^{\pm}\ell^{\pm}jb\missET$ sample.

\begin{figure}[t]
\includegraphics[width=0.48\linewidth]{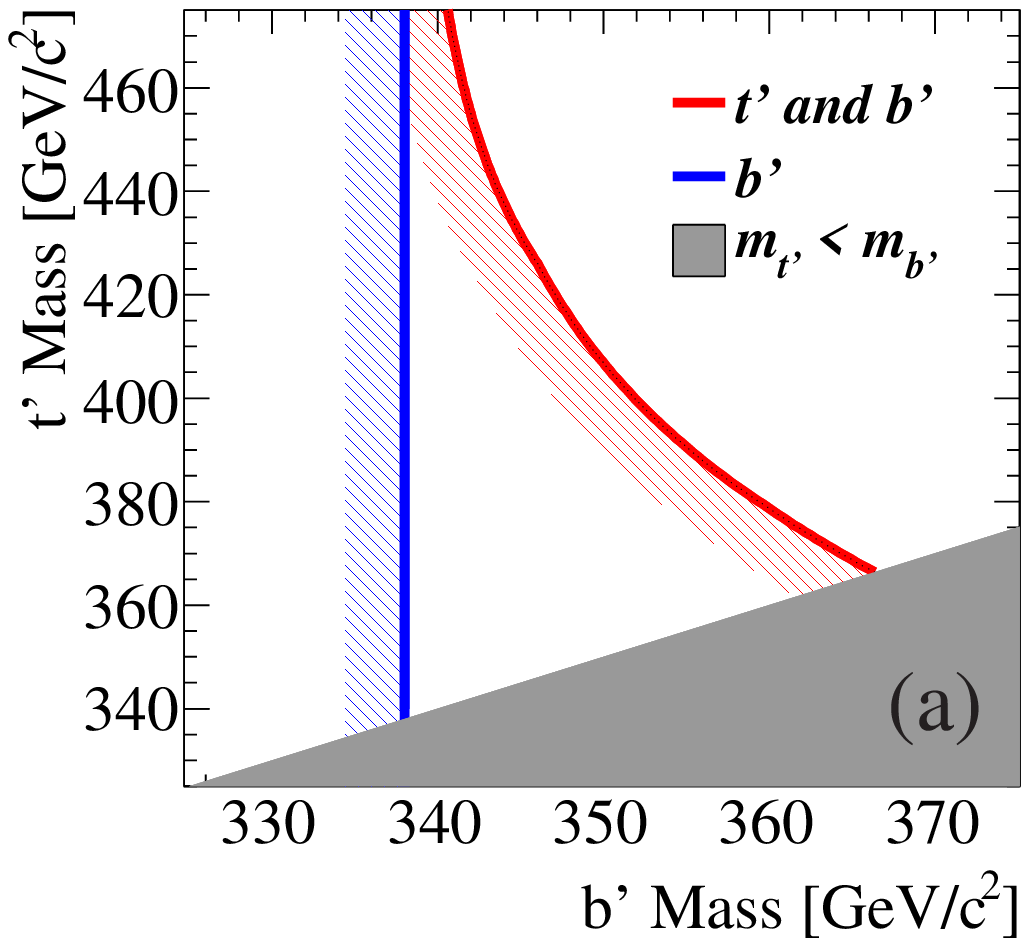}
\includegraphics[width=0.48\linewidth]{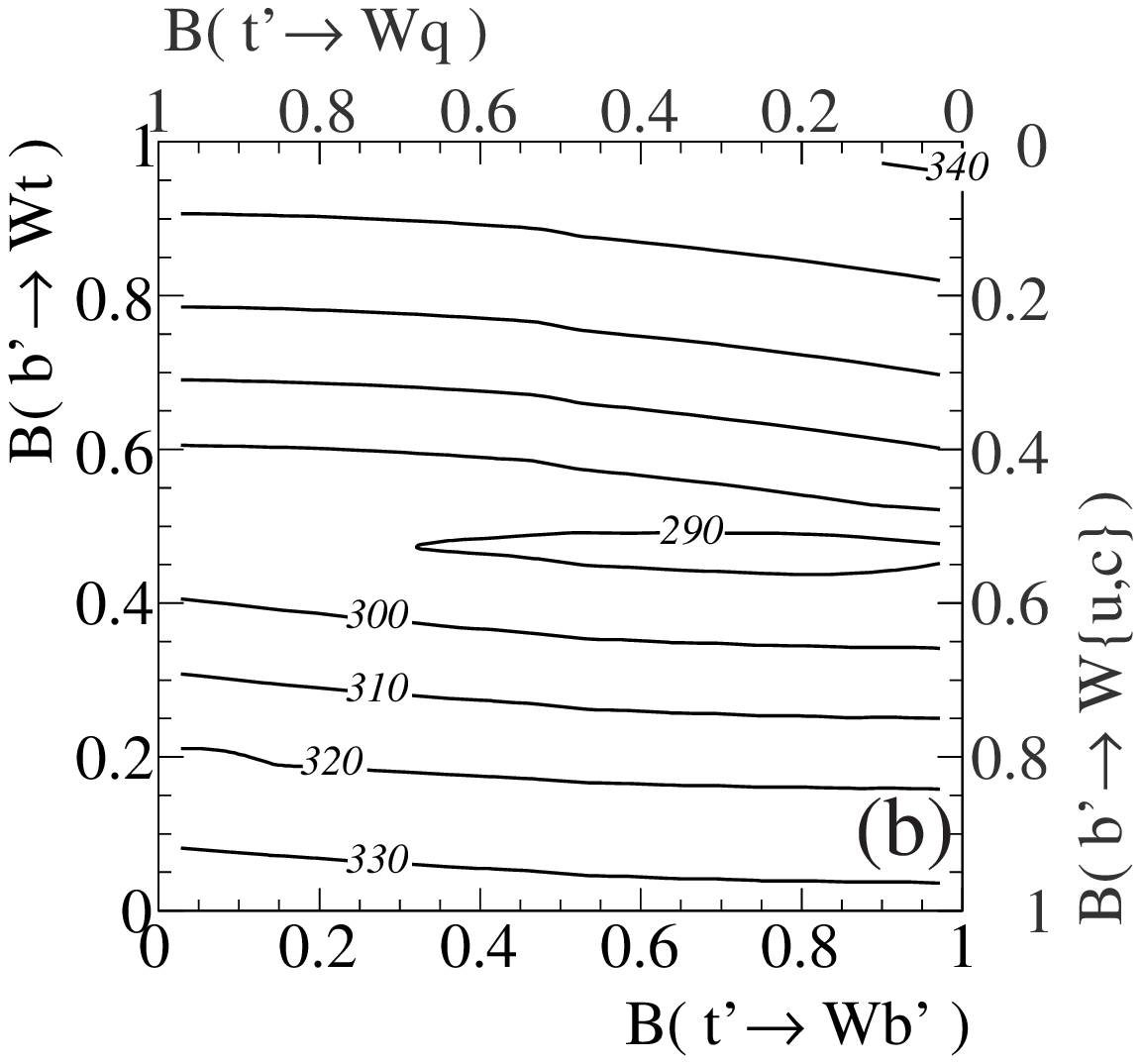} \\
\includegraphics[width=0.48\linewidth]{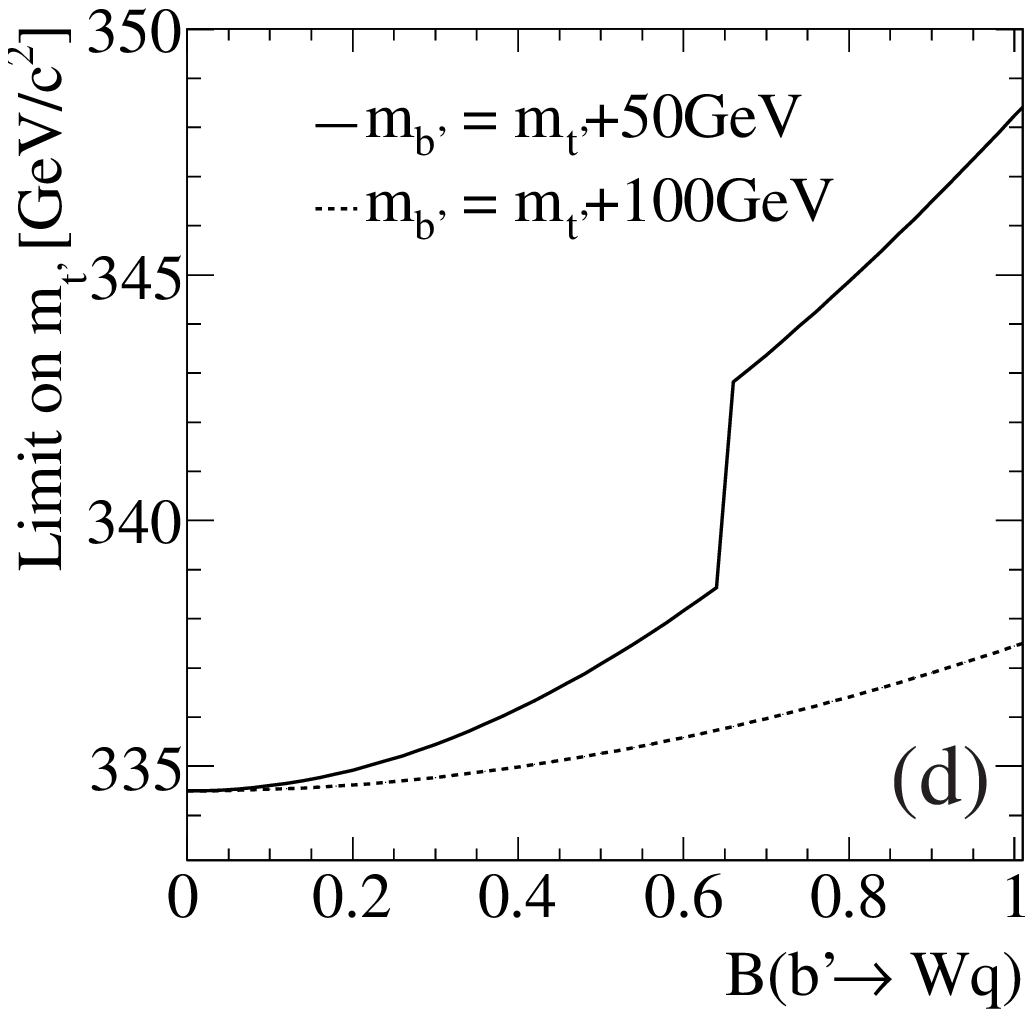}
\includegraphics[width=0.48\linewidth]{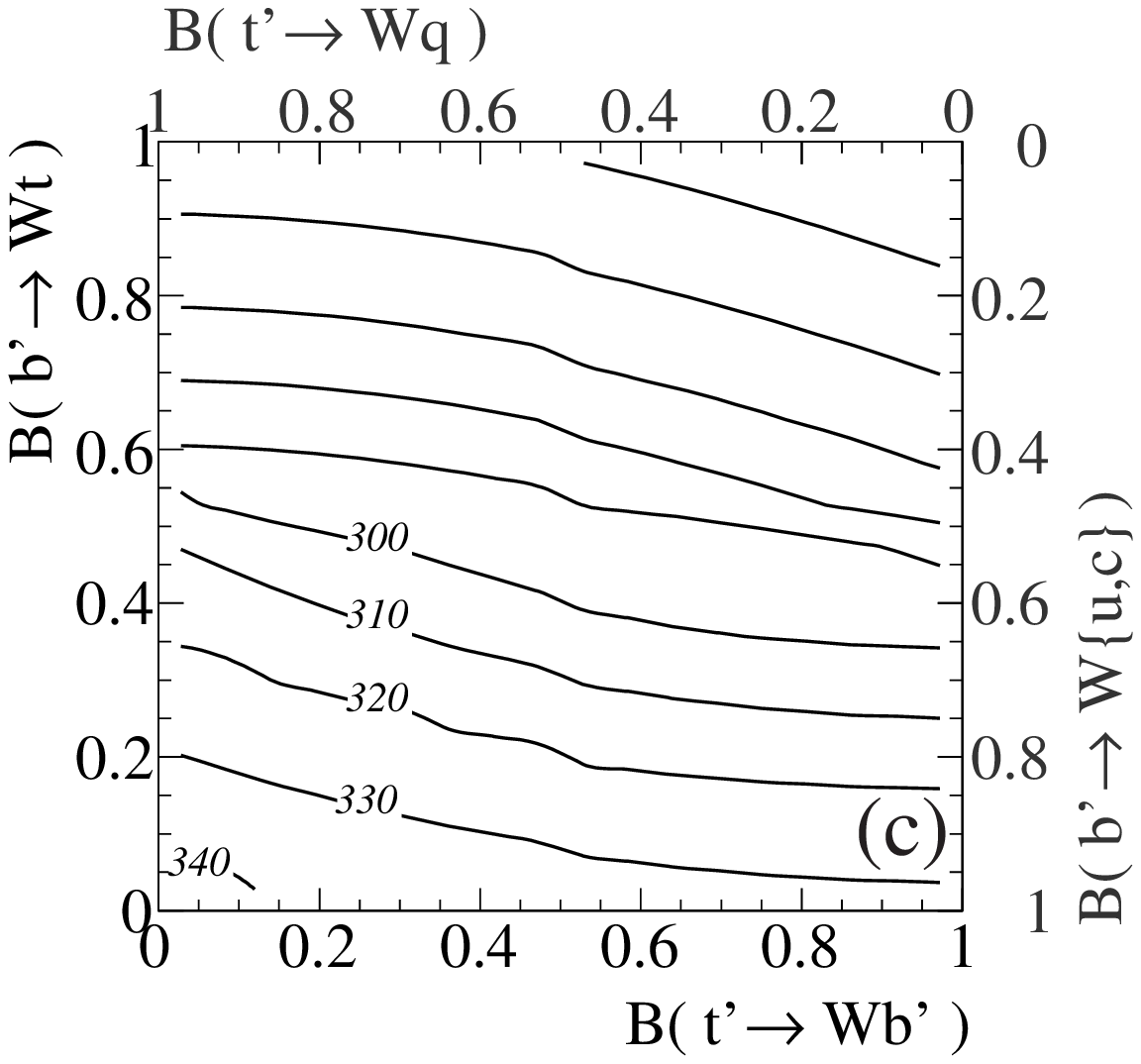}
\caption{(a) Limits for the combined hypothesis $t'\rightarrow Wb'$ and
  $b'\rightarrow Wt$ derived from the \lljbme data. Also plotted is the previous CDF limit for the individual $b^\prime$ case.
  (b) Limits on $b'$ mass from the combined $\ell^{\pm}\ell^{\pm}jb\missET$ and
  $\ell+4j$ data, as a function
  of branching fractions $\mathcal{B}(t'\rightarrow Wb') = 1 - \mathcal{B}(t'\rightarrow
  Wq)$ [$q=d,s,b$] and  $\mathcal{B}(b'\rightarrow Wt) = 1 - \mathcal{B}(b'\rightarrow
  Wq)$ [$q=u,c$] for the case $m_{t'} = m_{b'}+100~\textrm{GeV}$ (c) Same as in (b) but 
  for $m_{t'} = m_{b'}+50~\textrm{GeV}$. (d) Limits on $t'$ mass from $\ell+4j$ data as a function of
  BR$(b'\rightarrow W\{q=u,c\})$, in the inverted mass splitting case
  $m_{\bprime} > m_{\tprime}$.}
\label{fig}
\end{figure}

In general, the ratio of event yield to the integrated luminosity
$N/L$ equals the cross section multiplied by the acceptance rate. For
a particular process, such as an individual \bprime, this gives
the limit on cross section: $\sigma_{\bprime} = \frac{N} {L \cdot \epsilon_{\bprime}}$ where $\epsilon_{\bprime}$ is the acceptance rate for the observed process
within the experimental selection constraints. However, we can also consider
the case with two contributions (i.e., from \tprime and \bprime) if we know the relative 
acceptance rates between the corresponding
processes, $\epsilon_{rel}$, and if the two cross sections are dependent:
\begin{equation}
\frac{N} { (L\cdot\epsilon_{\bprime})} =  \sigma_{\bprime} + \epsilon_{rel}\cdot\sigma_{\tprime}(\footnotesize\sigma_{\bprime}).
\end{equation}

The \tprime and \bprime cross sections,
nearly the same up to the \tprime - \bprime mass difference, are both determined by perturbative QCD.
(Electroweak production
modes of pairs of the fourth generation quarks are expected to be $\leq 1\%$ at the
Tevatron). This is parameterized
from the next-to-leading order cross-section calculations for strong force production
of massive quarks
\cite{CSXmassive}, by fitting the published theoretical cross-section to an
exponential model in the local region of mass under consideration:
$\sigma_{\tprime}~=~\sigma_{\bprime}\exp{-\frac{m_{\tprime}-m_{\bprime}}{M}}$, where $M$ is found to be $40$~GeV.

This relationship yields cross-section limits and, by extension, mass limits
on \bprime for a spectrum of assumed \tprime
masses. The relative efficiency  $\epsilon_{rel}$ between the original model
and the model explicitly including both \tprime and \bprime was
estimated using simulated data.
The model for $\epsilon_{rel}$ accounts for increased acceptance of \tprime
as the fourth-generation mass difference
increases, with a plateau beyond $m_W$. This method was validated by
varying the model for $\epsilon_{rel}$. The resulting mass limits were found to be stable.
We find that the additional contribution
from the $t' \rightarrow Wb' $ decay gives the limits $m_{b'} > 340-360$ depending on
$m_{t'}$ (Fig.~\ref{fig}a), stronger than the previous CDF individual $b'$ limit.

We probe the full two-dimensional branching fraction interval
by calculating the dependence of the event yield on the branching fractions
explicitly. As the branching fractions to the reconstructed states vary, the acceptances and reconstruction
efficiencies of the
processes of interest vary accordingly.  Considering these effects, we calculate the acceptances of \bprime and \tprime
in both the \lfj and \lljbme samples.

The expected signal yield for one process relative to another is proportional to the relative
production rates and final-state reconstruction efficiencies. The relative signal production rates have two  factors:
the relative
cross section for initial state production (i.e., $t^\prime \bar t^\prime$ and $b^\prime \bar b^\prime$) and the
branching ratios of the involved processes. For several processes contributing to a signal
there are multiple terms of this form. With the event
yield fixed at its observed value, we isolate the cross section, and express it in terms of the previously
measured limit and an effective relative acceptance. The effective
 relative acceptance $(A)$ includes acceptance terms for all processes
considered, each scaled by the relative cross-section.

For the \lfj sample considered with the classical splitting case $m_{\tprime} > m_{\bprime}$, there are no relative reconstruction
efficiencies to consider so the expression for the relative expected yields as a function of $\beta_{b'} = \mathcal{B}(b'\rightarrow Wt)$ and
$\beta_{t'}=\mathcal{B}(t'\rightarrow Wb')$ is fairly simple:
\begin{equation}
 A(\beta_{b'},\beta_{t'}) = (1-\beta_{b'})^2 + \sigma_{rel}(1-\beta_{t'})^2
~,
\end{equation}
where $\sigma_{rel} \equiv \sigma_\tprime/ \sigma_\bprime$.
This treatment produces limits on the mass of the \bprime as a function of fourth-generation
branching fractions.

The \lljbme case is somewhat more complicated. In addition to \bprime, there are two significant \tprime
 processes that produce the signal selected for the sample analyzed:
 \begin{eqnarray}
  t'\rightarrow Wb' & \rightarrow & WWt \rightarrow WWWb ~, \nonumber \\
  t'\rightarrow Wb' & \rightarrow & WWc \nonumber ~.
 \end{eqnarray}
 
 Jet multiplicity and the flavor-tag requirement imply 
 different relative reconstruction efficiencies for each contribution. These factors are denoted
 $\epsilon_{NW}$, where $N$ is the number of intermediate $W$-bosons in the process described. They are estimated from
 simulated data and are calculated relative to the four-$W$ case considered in the original analysis.
The effective relative acceptance in this case has the form
\begin{eqnarray}
\lefteqn{A(\beta_{b'},\beta_{t'}) = \beta_{\bprime}^{2} + \beta_{\tprime}^{2}\frac{ \sigma_{rel}}{\epsilon_{bb}} [  (1-\beta_{b'})^2 \epsilon_{cc}} \nonumber \\
 & + & 2\beta_{b'}(1-\beta_{b'})\epsilon_{5W}\epsilon_{cb} + \beta_{b'}^2\epsilon_{6W}\epsilon_{bb}] ~,
\label{A2}
\end{eqnarray}
where the individual efficiencies for each combination of decay modes
are estimated by producing pairs of massive quarks at tree level
using the MadEvent software package \cite{Alwall:2007st} for a user-defined model containing $t'$ and $b'$
and allowing for ${\cal O} \sim 1\%$ mixing with either the third or second generations (i.e., $V_{4j} \sim {\cal O}(0.1)$ for $j=2,3$). Under
these assumptions, all of the inclusive decay widths (including $t' \rightarrow W b'$ which
is unsuppressed by mixing) are well within the expected jet energy resolution at a hadron
collider.  The events are generated at $b'$ masses of 400 GeV and for two different choices
of mass splitting, 50~GeV and 100~GeV, representing both on-shell and off-shell
$W$ bosons in the $t' \rightarrow b'$ decay.  The events are processed into the
desired decays modes using the BRIDGE program \cite{Meade:2007js}.
As expected, in the region of interest
the dependence of efficiencies on quark-mass choices is small.

Terms from the \tprime contribution have unique reconstruction efficiencies due to jet-flavor tagging, expressed in eq.~\ref{A2} by
$\epsilon_{f_1f_2}$, where $f_1f_2$ is the flavor combination of the quarks in the final state of the (quark-level) \tprime decay.
These efficiencies are given by statistics from the raw efficiencies of the jet-flavor tag to select the beauty (60\%)
or charm (15\%) flavor in a jet:
\begin{eqnarray}
\epsilon_{bb} & = & 1 - (1-\epsilon_b)^2 \nonumber \\
\epsilon_{cb} & = & \epsilon_c(1-\epsilon_b) +\epsilon_b(1-\epsilon_c) + \epsilon_b\epsilon_c \nonumber \\
\epsilon_{cc} & = & 1 - (1-\epsilon_c)^2 ~.
\end{eqnarray}

Using the above procedure, we produce limits on the mass of the \bprime as a function of fourth-generation
branching fractions.

The two sets of results reveal the complementary sensitivities of the CDF analyses
over the interval.
However, they are not orthogonal and thus cannot be statistically combined. Instead,
the best limits are found by choosing the stronger of two limits
at each point in the branching fraction interval.
The combined results are presented for two characteristic choices of mass splitting
in Fig.~\ref{fig}b,c.

Finally, we consider the case of an inverted splitting, $m_{\bprime} > m_{\tprime}$.
The CDF $\ell+4j$ data sets a limit on $t'$ for every choice of
branching fraction in this case, as its assumption that
$\mathcal{B}(t'\rightarrow Wq={d,s,b})=100$\% is always satisfied.  If
$\mathcal{B}(b'\rightarrow Wq={u,c}$) is significant it also contributes,
but at a smaller cross-section due to its larger mass. With the roles of
\bprime and \tprime reversed, $\sigma_{rel}$ is inverted accordingly in this case.
The relative acceptance as a function of $\beta_{b'} =\mathcal{B}(b'\rightarrow W t^\prime)$ is given by:
\begin{equation}
A(\beta_{b'}) = 1+ \sigma_{rel}\beta_{b'^2} ~.
\end{equation}

On the other hand The \lljbme sample, sensitive to $b$-type quarks, cannot at its present size
set limits in the inverted splitting case, as the limits it produces would imply
a lighter \tprime already excluded by the \lfj sample.

Our limits for the inverted case $m_{\bprime} > m_{\tprime}$ with the \lfj sample are shown in Fig.~\ref{fig}d. The displayed discontinuity in the upper curve of Fig.~\ref{fig}d is caused by the corresponding slope in the \lfj result.  


To summarize, by analyzing the expectations for relative event yields at CDF under 
the combined \tprime,\bprime hypothesis, and assuming a
continuum of \tprime and \bprime branching fractions, we find that the CDF data imply limits 
on $m_{\bprime}$ and $m_{\tprime}$ of $290$ GeV and greater over the full range of mixing scenarios, 
for two characteristic choices of the \tprime - \bprime mass splitting: 
$m_{\tprime} > m_{\bprime}$ and $m_{\bprime} > m_{\tprime}$.
The inclusion of a \tprime strengthens the previously obtained 
\bprime mass limit from the \lljbme sample; in
the $m_{\tprime} > m_{\bprime}$ case, by
up to 10\% when $0 < m_{\tprime} - m_{\bprime} < M_W$.

If a fourth generation of fermions is embedded
in theories beyond the SM, then the large splitting case
($m_{t'} - m_{b'} > M_W$)
and the inverted
scenario ($m_{b'} > m_{t'}$) may not be excluded.
Indeed, an example given recently in
\cite{hashimoto2} shows that precision EW data can accommodate
$m_{t'} - m_{b'} > M_W$
if there are two Higgs doublets.
In fact, the
compositeness picture emerging from the addition
of new heavy fermionic degrees of freedom is more
naturally described at low energies by multi-Higgs
theories \cite{hung,hashimoto1,hashimoto2},
for which constraints on the fourth generation parameter space
are known to be relaxed \cite{He:2001tp}.
In such cases (i.e., a large \tprime - \bprime mass splitting and/or 
an inverted splitting), we find new stronger bounds on 
$m_{b'}$ and $m_{t'}$, in some mixing scenarios reaching up to 
$430$ GeV. 

{\it Acknowledgments:} The work of CF and DW is supported in part by the Dept. of
Energy and the Alfred P. Sloan Foundation.  SBS acknowledge research support
from the Technion. TT is grateful to the SLAC theory group for their
 hospitality during which part of this work was completed.

\end{document}